\documentclass[%
aip,
amsmath,amssymb,
reprint,%
]{revtex4-1}
\usepackage{color}
\usepackage{natbib}
\usepackage{graphicx}
\usepackage{dcolumn}
\usepackage{bm}
\newcommand\tab[1][0.35cm]{\hspace*{#1}}
\usepackage[utf8]{inputenc}
\usepackage[T1]{fontenc}
\usepackage{mathptmx}
\usepackage{etoolbox}
\usepackage[colorlinks=true, urlcolor=blue, citecolor=blue, linkcolor= blue]{hyperref}
\makeatletter
\def\@email#1#2{%
	\endgroup
	\patchcmd{\titleblock@produce}
	{\frontmatter@RRAPformat}
	{\frontmatter@RRAPformat{\produce@RRAP{*#1\href{mailto:#2}{#2}}}\frontmatter@RRAPformat}
	{}{}
}%
\makeatother
\begin{document}
	
	\preprint{AIP/123-QED}
	
	\title{Endless Dirac nodal lines and high mobility in kagome semimetal  Ni$_{3}$In$_{2}$Se$_{2}$ single crystal\\}
	\author{Sanand Kumar Pradhan}
	\author{Sharadnarayan Pradhan}
	\affiliation{Department of Pure and Applied Physics, Guru Ghasidas Vishwavidyalaya, Koni, Bilaspur-495009, C. G., India.}
	\author{Priyanath Mal}
	\affiliation{Department of Physics and Photon Science, Gwangju Institute of
		Science and Technology (GIST), Gwangju 61005, Republic of Korea.}
			\author{P. Rambabu}
		\affiliation{Department of Pure and Applied Physics, Guru Ghasidas Vishwavidyalaya, Koni, Bilaspur-495009, C. G., India.}
	\author{Archana Lakhani}
	\affiliation{ UGC-DAE CSR, University Campus, Khandwa Road, Indore, 452001, India.}
	\author{Bipul Das}
	\affiliation{Department of Physics, National Taiwan Normal University, 162, Section 1, Heping E. Rd., Taipei City 106, Taiwan.}
		\author{Bheema Lingam Chittari}
		\affiliation{Department of Physical Sciences, Indian Institute of Science Education and Research Kolkata, Mohanpur 741246, West Bengal, India.}
	
			\author{G. R. Turpu}
		\affiliation{Department of Pure and Applied Physics, Guru Ghasidas Vishwavidyalaya, Koni, Bilaspur-495009, C. G., India.}

	\author{Pradip Das}
	\homepage[Electronic mail: ]{pradipd.iitb@gmail.com}
	\affiliation{Department of Pure and Applied Physics, Guru Ghasidas Vishwavidyalaya, Koni, Bilaspur-495009, C. G., India.}
	\

	\date{\today}
	
	\begin{abstract}
		Kagome-lattice crystal is crucial in quantum materials research, exhibiting unique transport properties due to its rich band structure and the presence of nodal lines and rings. Here, we investigate the electronic transport properties and perform first-principles calculations for  Ni$_{3}$In$_{2}$Se$_{2}$ kagome topological semimetal. First-principle calculations indicate six endless Dirac nodal lines and two nodal rings with a $\pi$-Berry phase  in the Ni$_{3}$In$_{2}$Se$_{2}$  compound. The temperature-dependent resistivity is dominated by two scattering mechanisms: $s$-$d$ interband scattering occurs below 50 K, while electron-phonon ($e$-$p$) scattering is observed above 50 K. The magnetoresistance (MR) curve aligns with the theory of extended Kohler's rule, suggesting multiple scattering origins and temperature-dependent carrier densities. A maximum MR of 120\% at 2 K and 9 T, with a maximum estimated mobility of approximately 3000 cm$^{2}$V$^{-1}$s$^{-1}$ are observed. The Ni atom's hole-like d$_{x^{2}-y^{2} }$ and electron-like d$_{z^{2}}$ orbitals exhibit peaks and valleys, forming a local indirect-type band gap near the Fermi level (E$_{F}$). This configuration enhances the motion of electrons and holes, resulting in high mobility and relatively high magnetoresistance.
		
	\end{abstract}
	
	\maketitle
	
		
		\begin{figure}
		\centering
		\includegraphics[width=0.9\linewidth]{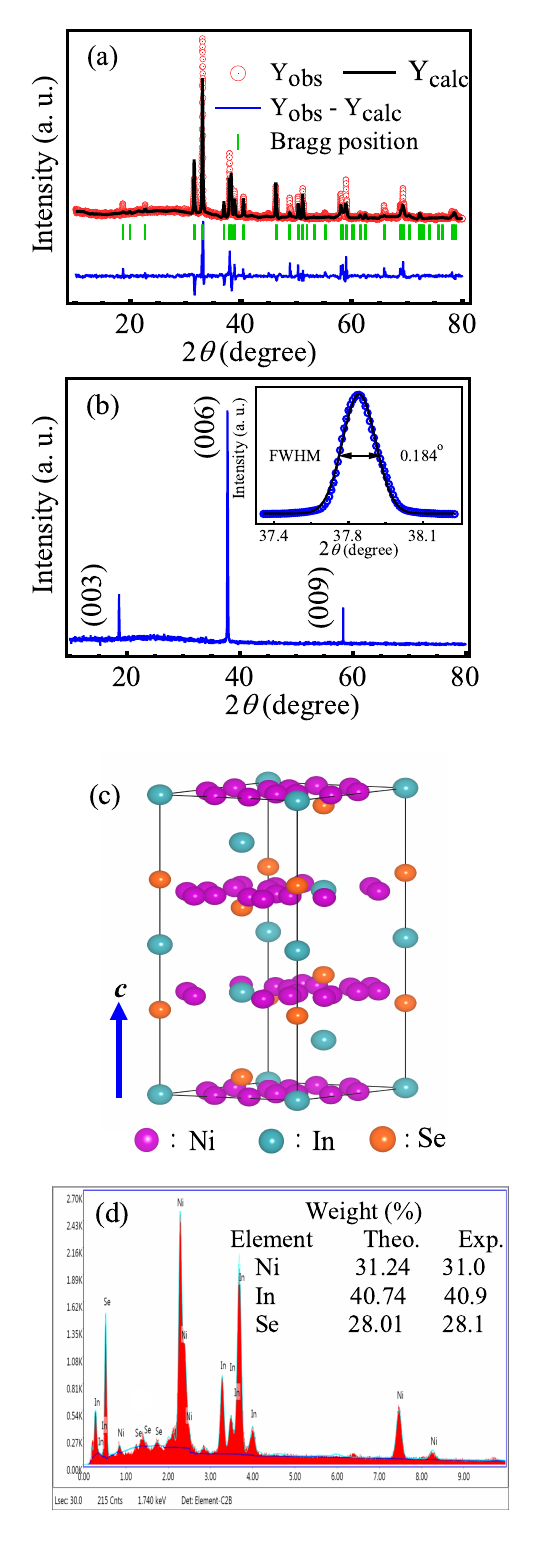}
		\caption{
			 (a) Rietveld refinement of room temperature polycrystalline x-ray diffraction pattern of Ni$_{3}$In$_{2}$Se$_{2}$ compound. (b) Single crystal x-ray diffraction pattern. (Inset figure illustrate FWHM corresponding \textit{(006)} plane.) (c) Unit cell of Ni$_{3}$In$_{2}$Se$_{2}$. (d) Energy dispersive spectroscopy of Ni$_{3}$In$_{2}$Se$_{2}$ single crystal.}
		\label{figure1}
	\end{figure}
	In recent years, kagome-lattice crystals, especially metals, have gained attention for potential applications in electronics and spintronics. Kagome metals, with a frustrated sub-lattice, explore  many exquisite quantum phenomena, including quantum Hall systems at low temperatures, strongly correlated materials, ferromagnetic quantum materials, and topological materials.\cite{a,b,c,d,e,f,g,h,j,i,mendels2016quantum} Theoretical models unveil nontrivial quantum interference, producing features like flat bands and Dirac fermions. Multi-band studies near the Fermi level reveal potential topological nodal lines, seen in ferromagnetic shandite kagome-metal Co$_{3}$Sn$_{2}$S$_{2}$,\cite{aa,bb,cc,dd} where the nodal ring transforms into Weyl nodes under strong spin-orbit coupling (SOC), with intrinsic Berry curvature contributing to a large anomalous Hall angle and giant anomalous Hall effect.\cite{ee,ff,gg,hh} A nonmagnetic counterpart shandite compound Ni$_{3}$In$_{2}$S$_{2}$ reveal a novel feature of endless Dirac nodal line by angle-resolved photoemission spectroscopy (ARPES) and density functional theory (DFT) calculations.\cite{zhang2022endless} In Ni$_{3}$In$_{2}$S$_{2}$, the SOC can be marginally increased by substituting Se at the S site in the kagome lattice, without affecting the magnetic contributions. According to recent literature, Ni$_{3}$In$_{2}$Se$_{2}$ is identified as a multi-band topological semimetal, showcasing a MR of 70\% and a mobility of 1000 cm$^{2}$V$^{-1}$s$^{-1}$.\cite{cao2023crystal} However, neither ARPES nor DFT calculations have reported the existence of endless Dirac nodal lines in this compound. Additionally, enhancing the mobility is crucial for the practical application of kagome materials in high-mobility electronic devices. \\
 	\tab Here, we present the endless Dirac nodal lines and nodal rings based on DFT calculations in a high quality kagome lattice of Ni$_{3}$In$_{2}$Se$_{2}$. Experimental electronic transport properties reveal the Ni$_{3}$In$_{2}$Se$_{2}$ is an electron-hole ($e$-$h$) compensated semimetal with high mobility  $\sim$3000 cm$^{2}$V$^{-1}$s$^{-1}$ and MR 120\% at 2 K. The electron-like d$_{x^{2}-y^{2} }$ and hole-like d$_{x^{2}-y^{2} }$ orbitals of Ni contribute to peaks and valleys in the electronic band structure, forming a local indirect type band gap near the Fermi level (E$_{F}$). This configuration accelerates the electron and hole motion and resulting in high mobility and MR. Two distinct scattering mechanisms, $e$-$p$ scattering above 50 K and $s$-$d$ interband scattering below 50 K, dominate the temperature-dependent resistivity. Unlike, the recent report where only one type of scattering mechanism is responsible for evolution of resistivity with temperatures.\cite{cao2023crystal} The MR curves follow extended Kohler's rule, suggesting a temperature-dependent carrier density variations rather than the ordinary Kohler's rule. \\
		\tab The single crystal of Ni$_{3}$In$_{2}$Se$_{2}$ was grown using a modified Bridgman technique. The crystal growth process involved using high-purity starting materials are Nickel powder (99.99\% purity), Indium pieces (99.99\% purity), and Selenium powder (99.99\% purity), which were double-sealed in an evacuated quartz tube under an argon atmosphere. The sealed tube was heated to 1323 K and slowly cooled to 773 K at a rate of 2 K/h. X-ray diffraction analysis for polycrystalline and single crystal samples was performed using Rigaku x-ray diffractometer with Cu-K$_{\alpha}$ $\lambda$ = 1.54 \AA. Energy dispersive spectra (EDS) were obtained using the FE-SEM model Quanta FEG 250 to assess nominal composition. Measurements of electronic transport, longitudinal resistivity, and Hall resistivity were performed on a freshly cleaved single crystal with typical dimensions of  2 mm $\times$ 1.3 mm $\times$ 0.21 mm. Measurements utilized conventional linear four-probe and five-probe techniques with the ACT options of the Physical Property Measurement System (PPMS) by Quantum Design Inc. USA.
		\begin{figure*}
		\centering
		\includegraphics[width=0.75\linewidth]{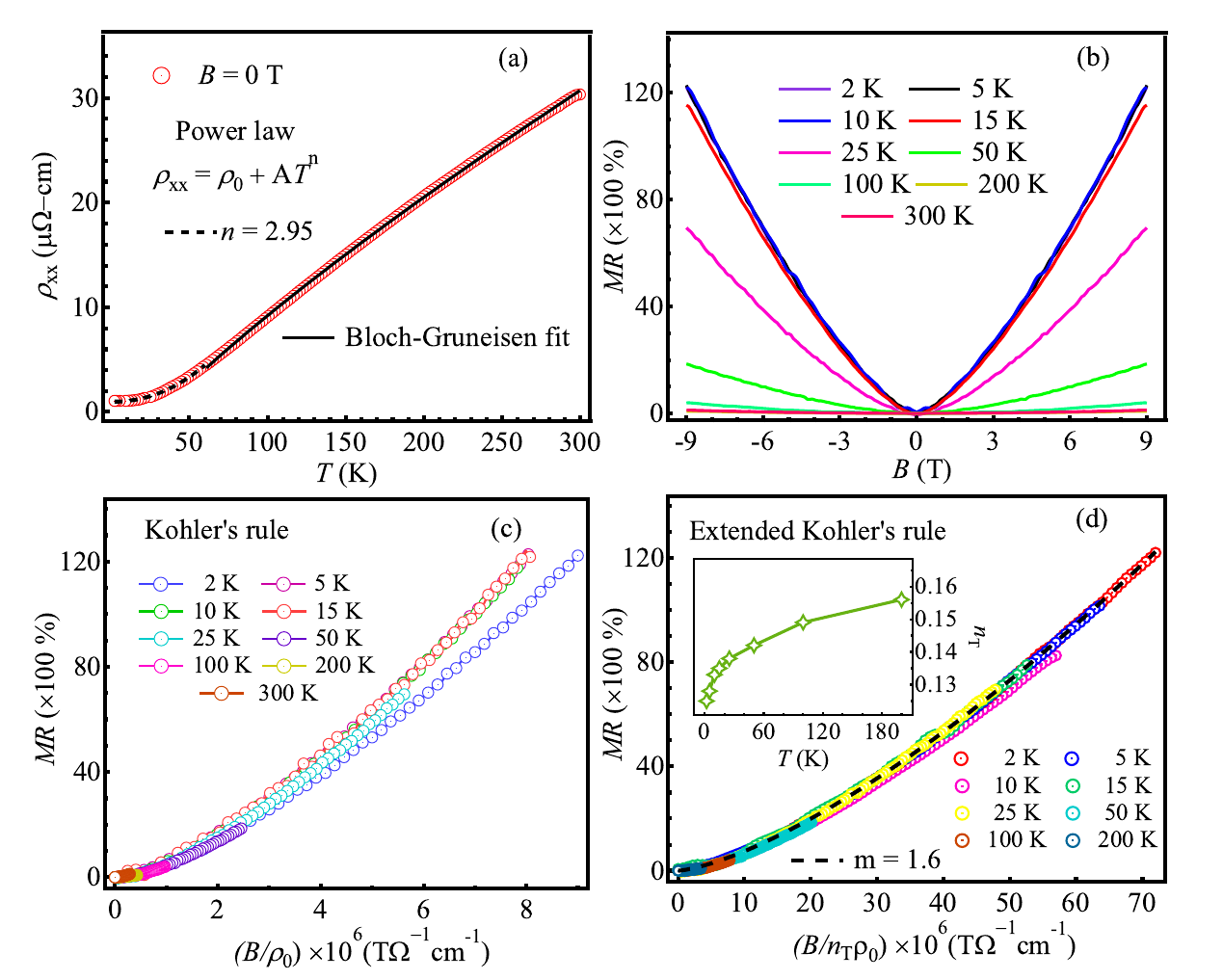}
		\caption{ (a) Longitudinal resistivity ($\rho_{xx}$) as a function of temperatures (\textit{T}) at zero magnetic field (\textit{B} = 0 T). The Bloch-Gruneisen fit (50 K to 300 K) is indicated by the solid black line, and the $s$-$d$ interband scattering ($n$ = 2.95) (2 K to 50 K) is represented by the dashed black line. (b) The variation of MR\%  with magnetic field at different temperatures.  (c) Illustrates violation of Kohler's rule analysis of the temperature dependence of the resistivity. (d) Illustrates validation of extended Kohler's rule analysis of the temperature dependence of the resistivity, and inset figure shows variation of density ($n$$_{T}$) with temperature.}
		\label{figure2}
	\end{figure*}
	\begin{figure*}
		\centering
		\includegraphics[width=0.75\linewidth]{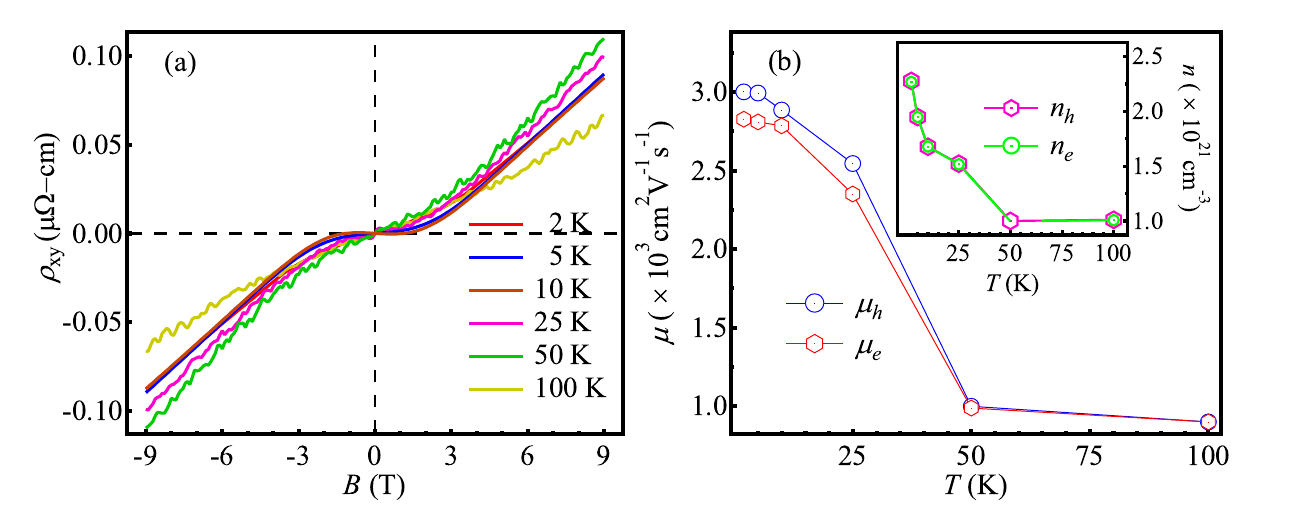}
		\caption{ Hall analysis of Ni$_{3}$In$_{2}$Se$_{2}$  single crystal. (a) Hall resistivity ($\rho_{xy}$) as a function of magnetic field with different temperatures. (b)  The carrier mobility of holes and electrons as a function of temperature  obtained from two-band model, whereas inset shows the carrier density of holes and electrons with variation of temperature.}
		\label{figure3}
	\end{figure*}
	\begin{figure*}
		\centering
		\includegraphics[width=1\linewidth]{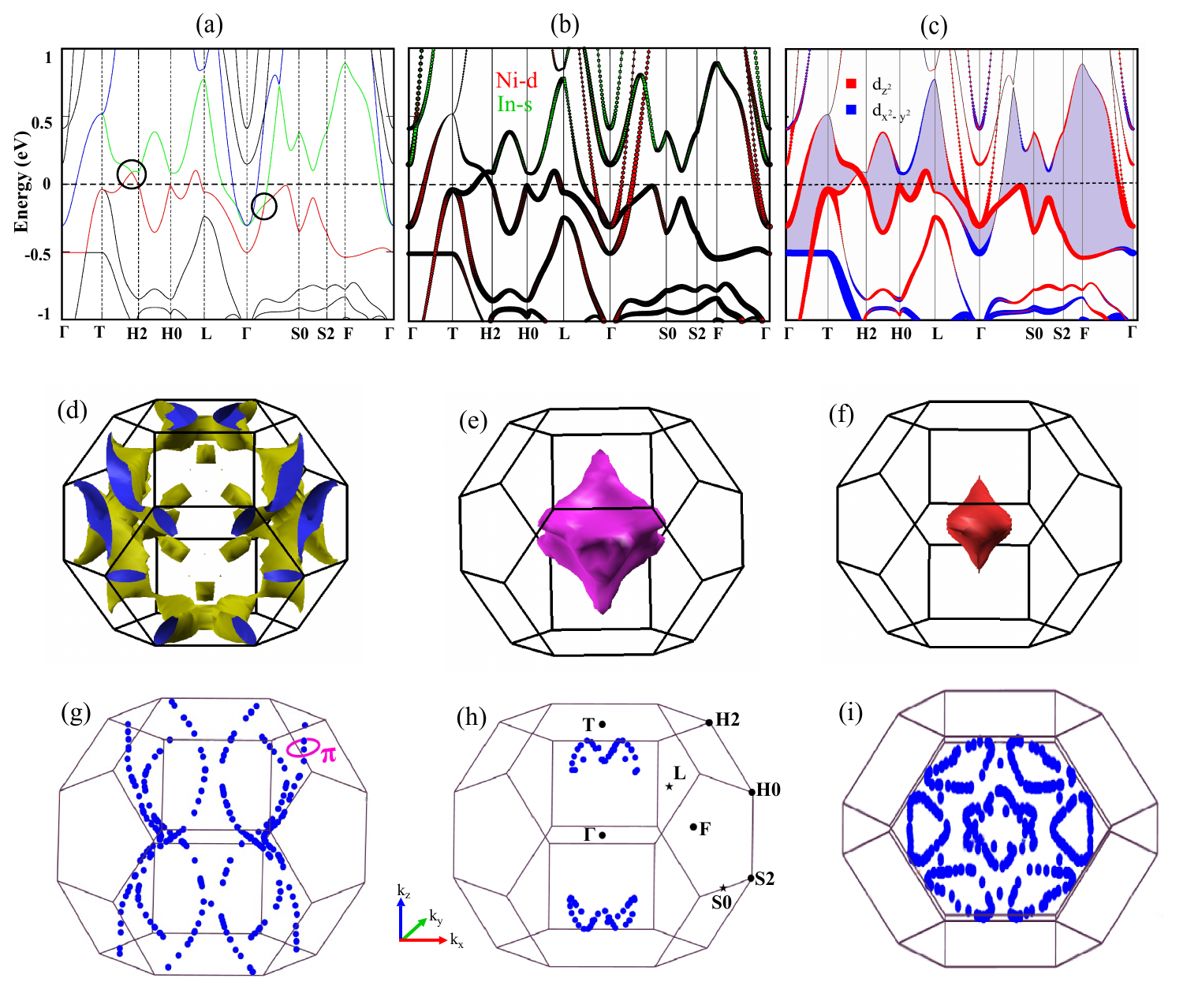}
		\caption{ (a) The electronic band structure of the Ni$_{3}$In$_{2}$Se$_{2}$ compound, without SOC; bands crossings near the Fermi level (E$_{F}$) are indicated by solid black circles.  (b) The character bands for Ni$_{3}$In$_{2}$Se$_{2}$  (the red and green colors correspond to Ni-d and In-s orbitals respectively. (c) Illustrates the band structure of Ni 3d orbitals in  Ni$_{3}$In$_{2}$Se$_{2}$, where the red and blue squares indicate the  d$_{z^{2}}$  and d$_{x^{2}-y^{2} }$ respectively. The shaded area indicates the local indirect-type band gap. (d), (e), and (f) display the Fermi surfaces of the Ni$_{3}$In$_{2}$Se$_{2}$  compound. (g)  Depiction of six endless nodal lines in blue, denoting band crossings in the first Brillouin zone, accompanied by a $\pi$-Berry phase.  (h) Exhibits two nodal rings within the Brillouin zone. (i) Provides a top-view representation of the nodal lines and nodal rings in the Brillouin zone for the Ni$_{3}$In$_{2}$Se$_{2}$ compound.}
		\label{figure4}
	\end{figure*}
	The electronic structure calculations were performed using the Vienna ab \textit{initio} simulation package (VASP) code.\cite{hohenberg1964inhomogeneous,kresse1996efficient,kresse1996efficiency} The generalized gradient approximations with Perdew-Burke-Ernzerhof (PBE) exchange-correlation functional was employed in the calculations.\cite{perdew1996generalized} A $\Gamma$-centered k-mesh of 15$\times$15$\times$15 was used for the sampling of the first Brillouin Zone (BZ). The crystal structure was optimized for a force convergence of 10$^{-3}$ eV\AA$^{-1}$, an electronic convergence of 10$^{-6}$ eV and a plane wave of cutoff energy of  516 eV respectively, in our calculations. The maximally localized wannier functions, as implemented in wannier90 code\cite{marzari1997maximally},  were utilized to construct the tight binding Hamiltonian.  The obtained tight binding Hamiltonian was further used in the wannier\_tools package\cite{marzari1997maximally,wu2018wanniertools}  to identify  topological nodal line features in the Ni$_{3}$In$_{2}$Se$_{2}$ compound.

Figure \ref{figure1}(a) shows the Rietveld refinements (RR) obtained from the x-ray diffraction patterns (XRD) of Ni$_{3}$In$_{2}$Se$_{2}$  powder samples. The analysis using Full-Prof software version 2019\cite{rodriguez2000introduction} reveals that the crystal structure of the Ni$_{3}$In$_{2}$Se$_{2}$ sample is in trigonal phase, belonging to the space group R$\bar{3}$m (\#166). The lattice parameters are determined to be $a$ = $b$ = 5.43 \AA, and $c$ = 14.24 \AA, resulting in a cell volume of 362.65 \AA$^{3}$. The XRD pattern for the mechanically exfoliated single crystal planes shows the preferred orientation along the crystallographic $c$-axis as shown in Fig. \ref{figure1}(b). We estimate the full width half maxima (FWHM) of the most intense peak \textit{(006)} of the single crystals, as shown in the inset of Fig. \ref{figure1}(b). The obtained FWHM value 0.184$^{o}$ indicates that the crystals is of excellent purity. The conventional cell of Ni$_{3}$In$_{2}$Se$_{2}$  is shown in Fig. \ref{figure1}(c), with the crystal structure sequence being In-Se-Ni-In-Se. The results of the EDS analysis are presented in Fig. \ref{figure1}(d), showing the weight percentages of the elements in the sample.\\
\tab Fig. \ref{figure2}(a) presents longitudinal resistivity (magnetic field, $B$ $\parallel$ $c$-axis \& current $I$ $\parallel$ $ab$ plane of the crystal) as a function of temperature in the range from 2 K to 300 K at $B$ = 0 T. The observed behavior is metallic-like. Ni$_{3}$In$_{2}$Se$_{2}$  single crystals exhibit a resistivity of 1.08 $\mu$$\Omega$-cm at zero field at 2 K, slightly higher than the previously reported value for Ni$_{3}$In$_{2}$S$_{2}$ single crystals.\cite{fang2023record} The residual resistivity ratio (RRR) of Ni$_{3}$In$_{2}$Se$_{2}$ single crystals, calculated as  ($\rho_{xx}$(300 K))/($\rho_{xx}$(2 K)), is determined to be 30, indicating high-quality of grown single crystals. Attempts are made to fit the temperature variation of the resistivity curve with the Bloch-Gruneisen (BG)\cite{ziman1960electrons} model between temperatures of 2 K and 300 K using the equation.
	\begin{equation}
	\rho_{xx}(T) =  \rho_{0} + A (\frac{T}{\theta_{D}})^5 \int_{0}^{(\frac{\theta_{D}}{T})} \frac{x^5}{(e^{x}-1)(1-e^{-x})}dx 
	\end{equation}
where $\rho_{0}$ is the residual resistivity $A$ is a constant and $\theta_{D}$ is the Debye temperature. The fitted values are $\rho_{0}$ = 0.478 $\mu$$\Omega$-cm and $\theta_{D}$ = 75 K respectively. The model fits well up to 50 K from 300 K, suggesting that in this temperature range, $e$-$p$ interactions control the scattering. $\rho_{xx}$(T) data below 50 K is inconsistent with the BG model; therefore, we fitted with a power law $\rho_{xx}$(T) = $\rho_{0}$ + $A$\textit{T}$^{n}$. The value of the $n$ is close to 3, indicating that the scattering mechanism is influenced by  phonon assisted $s$-$d$ interband scattering\cite{ziman1960electrons,hu2020high,wang2015origin,shekhar2015extremely}, therefore two different types of scattering mechanism are responsible for the evolution of temperature dependent resistivity in the range 2 K to 300 K.
The MR of Ni$_{3}$In$_{2}$Se$_{2}$ single crystals are studied, with $MR = \frac{\rho_{xx}(B)-\rho_{xx}(0)}{\rho_{xx}(0)}\times100\%$
where $\rho_{xx}$($B$) and $\rho_{xx}$(0) are the longitudinal resistivities at field $B$ and 0 T, respectively. 
Fig. \ref{figure2}(b) displays the MR variations with field $B$ at various temperatures. The maximum MR observed is 120\% at 2 K and 9 T. However, the MR decreases significantly as the temperature increases. We fit the MR data using Kohler's rule to better understand the role played by multiple scattering mechanisms and the temperature dependency of carriers. Kohler's rule states that in metals and topological or ordinary semimetals, the MR to an applied magnetic field takes the form of a functional parameter (magnetic field divided by residual resistivity), or MR=$f$($B$/$\rho_{0}$).\cite{kohler1938magnetischen,wang2015origin,narayanan2015linear} The main findings of the MR behavior in Ni$_{3}$In$_{2}$Se$_{2}$ are shown in Fig. \ref{figure2}(c) and Fig. \ref{figure2}(d), which demonstrate the violation of Kohler's rule and the validation of the extended Kohler's rule, respectively. The MR curves against $B$/$\rho_{0}$ in Fig. \ref{figure2}(c) do not collapse onto a single curve, indicating a clear violation of Kohler's rule. This is expected since the resistivity versus temperature data exhibit multiple scattering mechanisms, and it's possible that the carrier density varies with temperature. To incorporate temperature-dependent carrier density into Kohler's rule, we introduce a temperature-dependent multiplier1/$n_{T}$ to $B$/$\rho_{0}$ on the $x$-axis of the MR curve, known as the extended Kohler's rule.\cite{xu2021extended,wang2017large,sankar2017crystal,leahy2018nonsaturating,karn2023weak} Fig. \ref{figure2}(d) illustrates that Ni$_{3}$In$_{2}$Se$_{2}$ obeys the extended Kohler's rule. Temperature dependence of $n_{T}$ for single crystal Ni$_{3}$In$_{2}$Se$_{2}$ is shown in Fig. \ref{figure2}(d), which was obtained using an extended Kohler's rule\cite{xu2021extended} i.e., MR = $\alpha(B/n_{T}\rho_{0} )^{m}$; where $\alpha$ and $m$ are being constant parameters and their obtained values are $\alpha$ = 9.23$\times$10$^{-10}$ ($\Omega$cm/T)$^{1.6}$ and $m$ is 1.6 respectively.\\
\tab We measured the Hall resistivity ($\rho_{xy}$) at different temperatures between 2 K and 100 K as a function of magnetic field shown in Fig. \ref{figure3}(a). The non-linear behavior of the Hall resistivity indicates the existence of two different kinds of charge carriers in Ni$_{3}$In$_{2}$Se$_{2}$ single crystal. In order to examine this behavior, we used a two-band model to fit $\rho_{xy}$.\cite{shoenberg2009magnetic}
  \begin{equation}
 	\rho_{xy} = \frac{B}{e}\frac{(n_h\mu_{h}^2-n_e\mu_e^2)+(n_h-n_e)\mu_h^2\mu_e^2B^2}{(n_h\mu_{h}+n_e\mu_e)^2+(n_h-n_e)^2\mu_h^2\mu_e^2B^2}
 \end{equation}
  where $\mu_{h}$ and $\mu_{e}$  stand for hole and electron mobilities, respectively, and $n_{h}$ and $n_{e}$  denote the carrier densities of the holes and electrons, respectively. The hole and electron carrier densities at 2 K are, respectively,  $n_{h}$ $\sim$2.279(1)$\times$10$^{21}$ cm$^{-3}$ and $n_{e}$ $\sim$2.270(1)$\times$10$^{21}$ cm$^{-3}$, and the hole and electron mobilities are, respectively, $\mu_{h}$$\sim$2826 cm$^{2}$V$^{-1}$s$^{-1}$ and $\mu_{e}$$\sim$2999 cm$^{2}$V$^{-1}$s$^{-1}$. Between 2 K and 50 K, we saw a strong temperature dependence of $n_{h,e}$(T) and $\mu_{h,e}$(T), which might have resulted in a major violation of Kohler's rule in this temperature range.

 We now focus our attention onto the first-principles calculations. The conventional BZ  is shown in see Supplementary information (SI) Fig. S1(a). The optimized lattice constants are $a$ = $b$ =  5.44 \AA, and $c$ = 14.25 \AA, which are in close agreement with that of the experimental parameters. The dynamical stability of the optimized compound is ensured via positive phonon frequencies as shown in Fig. S1(b). The band structure of Ni$_{3}$In$_{2}$Se$_{2}$ without the inclusion of the SOC is shown in Fig. \ref{figure4}(a). It is seen from the Fig. \ref{figure4}(a) that there are three bands (shown in red, green and blue colors) that cross Fermi level E$_{F}$ indicating that the compound is semi-metallic. For the three crossed bands, we have plotted the corresponding three Fermi surfaces as shown in Fig. \ref{figure4}(d), \ref{figure4}(e), and \ref{figure4}(f).  The linear band crossings along T-H$_{2}$ (around 0.7 eV above E$_{F}$  ) and along $\Gamma$-S$_{0}$ ( around -0.156 eV below E$_{F}$ ) are shown in solid circles may give the nodal line behavior in this compound. These band crossings are mainly due to the Ni-d orbitals as can be seen from the character band plot as shown in Fig. \ref{figure4}(b). The three crossed bands give rise to various electron ($e$) and  hole ($h$) pockets. The red band has electron pocket nature along H$_{2}$-H$_{0}$ and H$_{0}$-$\Gamma$ and hole nature along the T-H$_{2}$  and H$_{0}$-T. The green band has  electron pocket nature along L-$\Gamma$ and hole nature along the H$_{2}$-H$_{0}$. The blue band has  electron pocket nature along $\Gamma$ and hole nature along the $\Gamma$-T. Altogether the entire electron and hole pockets gets compensated gives rise to $e$-$h$ compensation which is also confirmed from electronic transport measurements.\\
\tab The  d$_{z^{2}}$  and d$_{x^{2}-y^{2} }$ orbitals of Ni atom create a local indirect type band gap as shown in Fig. \ref{figure4}(c) around the Fermi level (E$_{F}$). The Ni-d$_{z^{2}}$  orbital possesses electron-like nature and d$_{x^{2} - y^{2} }$ has hole-like nature around E$_{F}$ (see in Fig. \ref{figure4}(c)) and have spherical Fermi surface like nature  due to valleys and peak nature. Due to this valley peak nature, the indirect band gap like behavior around  E$_{F}$, causes the motion of electrons and holes in a fast manner  resulting in high mobility and large MR in the Ni$_{3}$In$_{2}$Se$_{2}$  compound.
The nodal line features of Ni$_{3}$In$_{2}$Se$_{2}$  without SOC is shown  in 3D BZ as can be seen from Fig. \ref{figure4}(g). From Fig. \ref{figure4}(g) and \ref{figure4}(h), it is clear the Ni$_{3}$In$_{2}$Se$_{2}$ compound exhibits six endless Dirac nodal lines and two nodal rings  are observed.  These nodal lines and nodal rings possesses $\pi$-Berry phase as we found  that $\frac{1}{2\pi} \oint_{s} F_{n}(k).dS $= 1;\cite{gangaraj2017berry} similar kind of endless Dirac nodal lines were observed recently in  the isostructural compound Ni$_{3}$In$_{2}$S$_{2}$.\cite{zhang2022endless} The band structure of Ni$_{3}$In$_{2}$Se$_{2}$ with SOC is shown in Fig. S1(c). From Fig. S1(c), it is clearly seen that the nodal line features are destroyed with the application of SOC.\\
      \tab In summary, we have grown high-quality kagome crystals of Ni$_{3}$In$_{2}$Se$_{2}$, exhibiting compensated semi-metallic behavior with two distinct scattering mechanisms in the temperature-dependent resistivity data: $s$-$d$ interband scattering below 50 K and electron-phonon scattering above 50 K. The carrier densities vary with temperature, aligning with the extended Kohler's rule. Magnetoresistance measurements yield 120\% MR, while Hall measurements estimate carrier mobility of $\sim$3000 cm$^{2}$V$^{-1}$s$^{-1}$. DFT calculations reveal endless Dirac nodal lines and nodal rings with a $\pi$-Berry phase in the Ni$_{3}$In$_{2}$Se$_{2}$ crystal. The electronic band structure displays a local indirect-type band gap close to the Fermi level (E$_{F}$), formed by peaks and valleys of the hole-like d$_{x^{2}-y^{2} }$ and electron-like d$_{z^{2}}$ orbitals of the Ni atom. The observed high carrier mobilities and comparatively high MR result from electrons and holes moving faster through these peaks and valleys.

		\begin{acknowledgments}
P. Das acknowledge UGC-DAE CSR, Indore center for the financial support through the project Ref. No. CRS/2021-22/01/425. The authors of GGV are acknowledging DST, Government of India for supporting the Department of Pure and Applied Physics through FIST Level-I program. We also acknowledges UGC-DAE CSR, Indore for electronic transport measurements using the PPMS.
		\end{acknowledgments}
		\section*{Conflict of Interest}
			The authors have no conflicts to disclose.
		\section*{Data Availability Statement}
		The data are available from the corresponding author upon reasonable request.

		\nocite{*}
		\bibliography{references}

	\end{document}